# Monte Carlo study of phase transitions in model orthonickelate

V. S. Ryumshin[1]

[1]Institute of Natural Sciences and Mathematics, Ural Federal University, Ekaterinburg, Russia

E-mail: vitaliy.riumshin@urfu.ru

**Abstract**

The results of numerical simulation using a classical Monte Carlo method with a kinematic accounting of the bosons concentration for a pseudospin model of orthonickelates are presented. Type of the phase transitions of the model orthonickelates is investigated.

**Keywords:** orthonickelates, Monte-Carlo, phase transitions

# 1 Introduction

Orthonickelates RNiO$_3$ (R – rare earth or Y) possess a number of unusual and interesting properties, primarily a metal-insulator transition, unusual conductivity behavior, and the presence of non-collinear magnetic structures. The main topic of discussion in nickelates is the phase transitions to insulating and magnetic states, which are also accompanied by structural changes in the system. A definitive hierarchy between the electronic, magnetic, and structural effects that occur during phase transitions in orthonickelates has yet to be established [1-4].

In this paper, model orthonickelate is considered as a charge-disproportionated phase described by a spin-triplet boson system with a quartet of states at a site. Previously, phase states within this model were investigated and phase diagrams were obtained for a number of parameters using the mean-field approximation and classical Monte Carlo numerical simulations [5,6]. This work represents a continuation of the study of model nickelate, with an emphasis on phase transitions, in particular, determining the transition types.

# 2 Monte Carlo method and Binder cumulants

For simulations used classical Monte Carlo method with a kinematic accounting of the bosons concentration. This algorithm was discussed in detail in the work [5].

Type of phase transitions was determined by Binder cumulant method [7]. The fourth-order Binder cumulants have the following form:

$$V_L = 1 - \frac{\langle E^4 \rangle_L}{3 \langle E^2 \rangle_L^2},$$

$$U_L = 1 - \frac{\langle O^4 \rangle_L}{3 \langle O^2 \rangle_L^2},$$

where $V_L$ – energy cumulant, $U_L$ – order parameter cumulant (magnetic and charge). This method is also used for high-precision determination of critical temperatures for various types of phase transitions.

# 3 Results

Figures 1 and 2 show the dependences for the energy cumulant during the NO-AFM transition and the charge cumulant during the NO-CO transition, respectively. For a second-order phase transition, the $V_L$ value tends to 2/3 at all temperatures, and $U_L$ will intersect for various system sizes at a point corresponding to the critical temperature. For a first-order transition, $V_L$ tends to 2/3 at high and low temperatures, and at the critical temperature, it tends to a value different from 2/3.

Both cumulants predict a second-order phase transition for both cases in a two-dimensional system. This is also confirmed by the energy histograms. A similar result is observed in the system of singlet hard-core bosons.

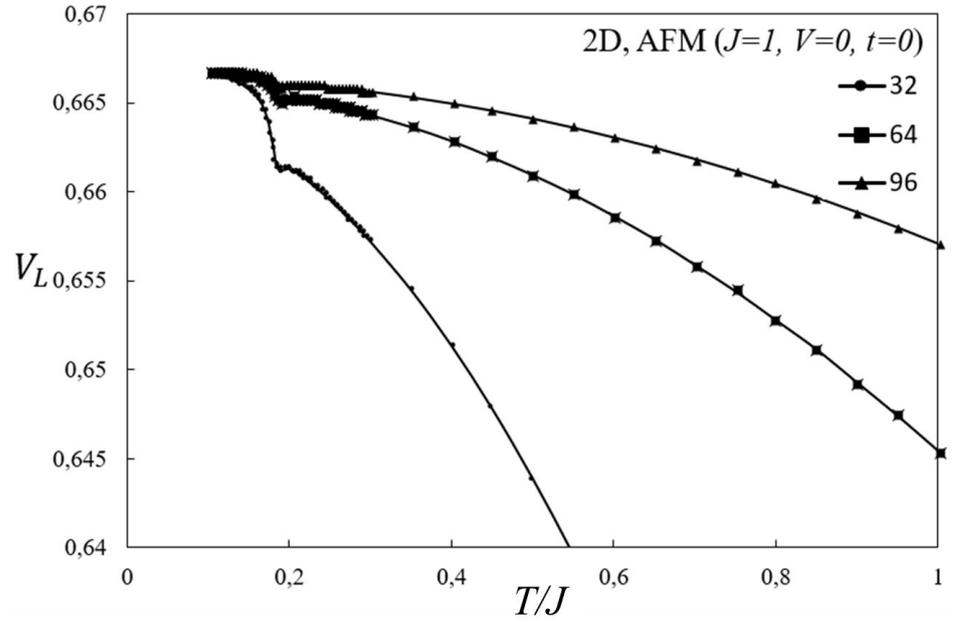

Figure 1. Energy cumulant $V_L$ of NO-AFM transition. Markers correspond to different grid sizes (L).

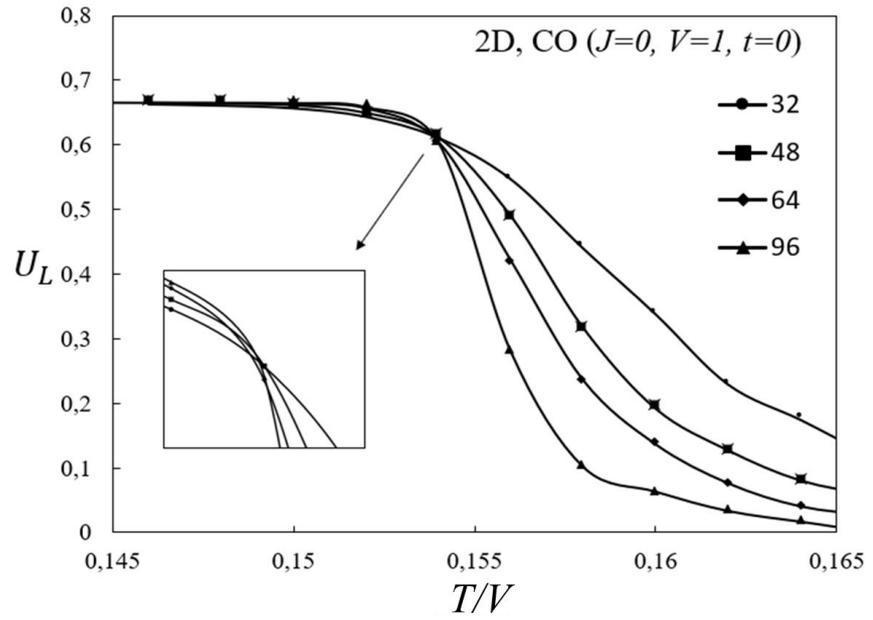

Figure 2. Charge cumulant $U_L$ of NO-CO transition. Markers correspond to different grid sizes (L).

However, in the case of a three-dimensional system, a first-order phase transition is observed during the NO-CO transition. This is demonstrated by the energy cumulant (Fig. 3), as well as the energy distribution histograms $P(E)$ for L = 8 (Fig. 4) and L = 32 (Fig. 5), respectively. Furthermore, the sharp change in energy and order parameter upon transition to the CO phase supports a first-order transition.

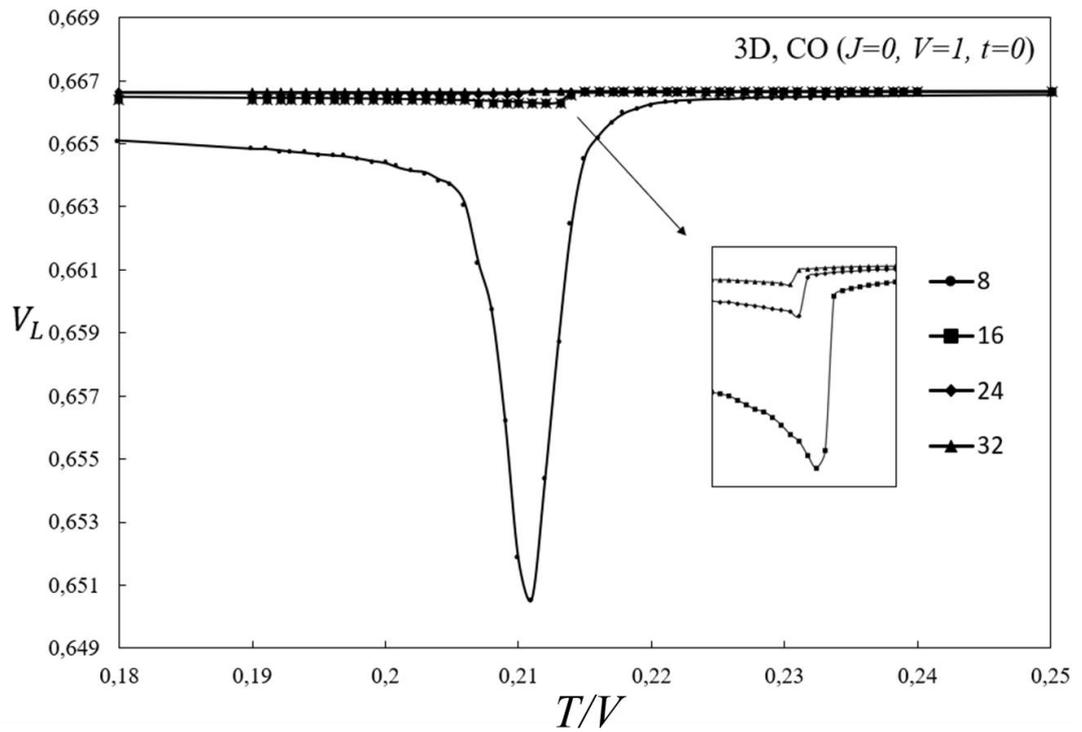

Figure 3. Energy cumulant $V_L$ of NO-CO transition. Markers correspond to different grid sizes (L).

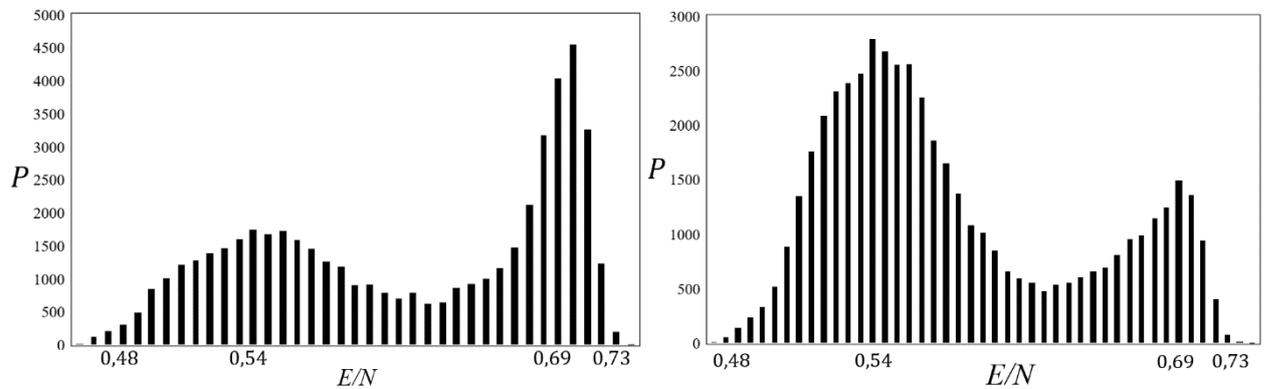

Figure 4. Energy histogram of NO-CO transition, size system is 8x8x8.

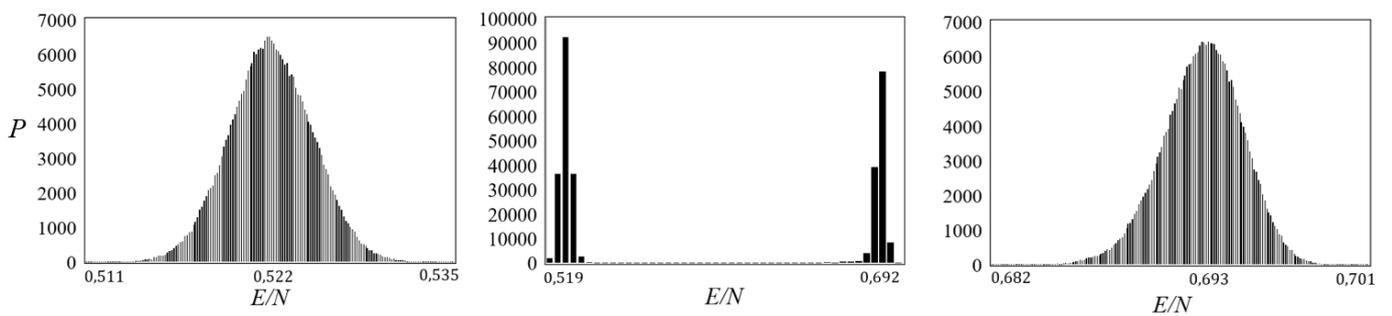

Figure 5. Energy histogram of NO-CO transition, size system is 32x32x32.

# 4 Conclusion

The results of numerical simulation using the classical Monte Carlo method with a kinematic accounting of the bosons concentration predict second order phase transition for NO-CO and NO-AFM transitions for 2D system and first order phase transition for NO-CO transition for 3D system.

## Funding

This study was supported by the Ministry of Science and Higher Education of the Russian Federation, project FEUZ-2023-0017.

## References

1. D. J. Gawryluk, Y. M. Klein, T. Shang, D. Sheptyakov, L. Keller, N. Casati, P. Lacorre, M. T. Fernandiez-Diaz, J. Rodriguez-Carvajal, and M. Medarde. Physical Review B **100**, *20*, 205137 (2019).

2. S. Catalano, M. Gibert, J. Fowlie, J. Íñiguez, J. M. Triscone, and J. Kreisel. Reports on Progress in Physics **81**, *4*, 046501 (2018).

3. M. L. Medarde. J. Phys. Condens. Matter **9**, *8*, 1679 (1997).

4. Y. M. Klein, M. Kozlowski, A. Linden, Ph. Lacorre, M. Medarde, and D. J. Gawryluk. Crystal Growth & Design **21**, *7*, 4230 (2021).

5. Ю. Д. Панов, С. В. Нужин, В. С. Рюмшин, А. С. Москвин. Физика твердого тела **66**, *7*, 1221 (2024).

6. В. С. Рюмшин, С. В. Нужин, Ю. Д. Панов, А. С. Москвин. Физика твердого тела **66**, *7*, 1047 (2024).

7. A. M. Ferrenberg, R. H. Swendsen. Physical review letters **61**, *23*, 2635 (1988).